\documentclass[journal]{IEEEtran}
\usepackage[utf8]{inputenc}
\usepackage{balance}
\usepackage{booktabs}
\usepackage[T1]{fontenc}
\usepackage{physics,amsmath}
\usepackage{amsthm}
\usepackage{mathtools}
\usepackage{amsfonts}
\usepackage{amssymb}
\usepackage{graphicx}
\usepackage{makeidx}
\usepackage{csquotes}
\usepackage{cite}
\usepackage{xcolor}
\usepackage{subcaption}
\usepackage{float}
\usepackage{multicol}
\usepackage{multirow}
\usepackage{array}
\usepackage{pifont}
\usepackage{hyperref}
\usepackage[capitalise,nameinlink ,noabbrev]{cleveref}

\usepackage[utf8]{inputenc}
\usepackage[algo2e,ruled, vlined, linesnumbered]{algorithm2e}

\usepackage[justification=justified,singlelinecheck=false]{caption}
\usepackage{caption}
\usepackage[hyphenbreaks]{breakurl}
\begin{document}
	\title{\LARGE UAV-Assisted 5G Networks: Mobility-Aware 3D Trajectory Optimization and Resource Allocation for Dynamic Environments  }
\author{Asad Mahmood, Thang X. Vu, Wali Ullah Khan, Symeon Chatzinotas, Bj\"orn Ottersten \\Interdisciplinary Centre for Security, Reliability and Trust (SnT), University of Luxembourg\\
\{asad.mahmood, thang.vu,waliullah.khan, symeon.chatzinotas, bjorn.ottersten\}@uni.lu  }%
	\markboth{IEEE VTC2025-Fall}%
{Shell \MakeLowercase{\textit{et al.}}: Bare Demo of IEEEtran.cls for IEEE Journals} 
\maketitle 
\begin{abstract}
This work proposes a framework for the robust design of UAV-assisted wireless networks that combine 3D trajectory optimization with user mobility prediction to address dynamic resource allocation challenges. We proposed a sparse second-order prediction model for real-time user tracking coupled with heuristic user clustering to balance service quality and computational complexity. The joint optimization problem is formulated to maximize the minimum rate. It is then decomposed into user association, 3D trajectory design, and resource allocation subproblems, which are solved iteratively via successive convex approximation (SCA). Extensive simulations demonstrate: (1) near-optimal performance with $\epsilon \approx 0.67\%$ deviation from upper-bound solutions, (2) $16\%$ higher minimum rates for distant users compared to non-predictive 3D designs, and (3) $10-30\%$ faster outage mitigation than time-division benchmarks. The framework's adaptive speed control enables precise mobile user tracking while maintaining energy efficiency under constrained flight time. Results demonstrate superior robustness in edge-coverage scenarios, making it particularly suitable for $5G/6G$ networks.
\end{abstract}
\begin{IEEEkeywords}
UAV-assisted communication, 3D trajectory optimization, User mobility prediction,  Successive Convex Approximation (SCA), Dynamic resource allocation
\end{IEEEkeywords}
\vspace{-2mm}
\section{Introduction} 
Recent advancements in 6G communication systems enable applications that demand massive connectivity, ultra-reliable low-latency communication, and high data rates \cite{10422956}. However, terrestrial base stations (BS) remain vulnerable to disruptions from natural disasters or network congestion \cite{10122889}. 
An attractive alternative, unmanned aerial vehicles (UAVs) have emerged as a flexible alternative, leveraging mobility and reliable line-of-sight (LoS) links to act as flying base stations (BS), mobile relays, or data collectors \cite{8247211}. Besides these advantages, UAVs face challenges, including finite 
 battery life \cite{9714909} that limits flight time by $16\%$\cite{7918510}, security risks, regulatory barriers, and communication resources. Moreover, in practical scenarios, UAVs lack real-time user location \cite{10123705}, further bottlenecks in resource allocation and trajectory planning, particularly for mobile users requiring seamless connectivity. Additionally, higher UAV altitudes improve LoS but increase path loss, thus reducing communication efficiency \cite{10122731}. To address these challenges, ongoing research focuses on UAV trajectory/placement optimization, real-time user mobility prediction, and dynamic resource allocation, ensuring seamless connectivity and enhanced QoS in next-generation networks.\par
Recent work on UAV-assisted networks focuses on 2D \cite{9714909,9798245,10122889} and 3D deployment \cite{10122731} to improve system resource usage. Static deployment, however, fails to adapt to dynamic environments. Researchers now emphasize trajectory design, allowing UAVs to adjust positions dynamically for greater efficiency. Early studies optimized 2D trajectories at fixed altitudes \cite{10529184,8247211},  whereas recent works focus on 3D design \cite{10422956,10100674,10181121}  which exploit altitude variations for performance enhancement. For example, \cite{10422956} uses deep reinforcement learning to optimize multi-UAV coverage and energy efficiency under known user distributions. In \cite{10100674} authors balances energy harvesting and data collection in solar-powered UAVs via max-min fair 3D trajectory design, while \cite{10181121} models 3D energy consumption under stochastic wind, proposing adaptive trajectories for energy-efficient communications. Critically, these studies assume static users—an unrealistic premise in dynamic environments where user mobility directly impacts network performance and UAV path planning.\par
To address this, several studies have been conducted on mobile users, where the user mobility model follows the Reference Point Group Mobility (RPGM) model \cite{10123705,10176326}, the Gauss-Markov Mobility (GMM) model \cite{9757149,9473504}  or real-world mobility data collected from Twitter through the Twitter API \cite{8727504}. These models have been applied to both 2D trajectory design \cite{10123705,10176326,9678115,9473504} and 3D trajectory design \cite{9757149,8727504} under various assumptions and objectives. Moreover, the studies \cite{10123705,10176326,9678115,9757149,9473504} assume that UAVs have access to users' locations at each time step through Global Positioning System (GPS), wireless sensing, or feedback, among other methods. In contrast, continuous tracking through feedback/signaling results in computational overhead, accelerating battery depletion and reducing UAV flight duration. Similarly, the authors of \cite{10123705,9757149,9678115,10176326}  consider a simplified LoS model and do not explicitly account for NLoS components. Moreover, the authors \cite{9757149,9473504} optimize the trajectory design under the assumption of fixed communication resources, which limits the system's flexibility. The approach in \cite{8727504} uses an Echo State Network (ESN) to predict user locations but neglects practical constraints like maximum speed, acceleration, and user connectivity time. Optimizing connectivity and flying time is crucial for improving UAV efficiency and system performance. \textit{This work overcomes prior limitations by implementing a dynamic 3D trajectory design with a predictive model for user location estimation. It optimizes flight time and adapts service time allocation while dynamically adjusting resources to enhance efficiency and responsiveness.} The main contributions of this work are listed below:
\subsection{Contribution}
We propose a 3D UAV trajectory design framework integrated with a Markov chain-based user mobility predictor, user association, and dynamic resource allocation to maximize the minimum rate in congested/disaster scenarios. The NP-hard joint optimization problem is decomposed via Block Coordinate Descent (BCD) into two subproblems: (i) user association and 3D trajectory optimization and (ii) communication resource allocation, solved iteratively using predicted user locations. To reduce complexity, a heuristic clustering method groups users by location, demand, and ergodic capacity, enabling tractable trajectory and resource allocation via Successive Convex Approximation (SCA) with guaranteed (theoretical) convergence and polynomial-time complexity. Simulations demonstrate near-optimal performance (0.67\% deviation), 15.51\% higher rates for edge users versus non-predictive 3D designs, and 10–30\% faster outage mitigation than time-division benchmarks, with adaptive speed control ensuring energy-efficient tracking in dynamic environments.
\section{System Model}
\label{SM}
\allowdisplaybreaks
We consider a UAV-assisted downlink wireless network for intermittent coverage to $N$ mobile users, indexed by $n \in \mathcal{N} = \{1, \dots, N\}$ within a specified area. A UAV, acting as a flying base station (BS) with $M$ antennas, operates for up to $T$ time slots and serves users using orthogonal frequency bands per time slot. Given the UAV's limited prior knowledge of user positions, a Gauss-Markov mobility model \cite{9757149,9473504} generates synthetic (historical) data to train a prediction model at the Central Control Unit (CCU). The CCU estimates user locations based on the historical data and transmits this information to the UAV via a dedicated control link before flight. To ensure efficient service, the UAV clusters users into $L$ groups \cite{10122731}, each containing $N_l$ users ($\mathcal{N}_l \subset \mathcal{N}$). The service duration for cluster $l$ is $T_l$, where $T_l<T$ and $\sum_{l \in \mathcal{L}} T_l \le T$. Each $T_l$ is further divided into discrete time slots of $\delta$ seconds, assuming stationary users within a slot \cite{8247211}. At time slot $t_l \in \mathcal{T}_l$, the estimated position of user $n$ in cluster $l$ is $\boldsymbol{Q}^l_n[t_l] = [x^l_n[t_l] + \Delta x^l_n, y^l_n[t_l] + \Delta y^l_n]^T$, where $x^l_n[t_l]$ and $y^l_n[t_l]$ are actual positions, and $\Delta x^l_n, \Delta y^l_n$ denote prediction errors. The UAV’s 3D position is $\boldsymbol{W}[t_l] = [\boldsymbol{w}^T[t_l], H[t_l]]^T$, where $\boldsymbol{w}[t_l] = [x[t_l], y[t_l]]^T$ represents horizontal coordinates and $H[t_l]$ the flight altitude. The UAV's trajectory is constrained by maximum speeds for horizontal ($V_{xy_{\text{max}}}$) and vertical ($V_{H_{\text{max}}}$) movements.
\begin{align}
&x_{\text{min}} \le x[t_l] \le x_{\text{max}}, \quad \forall l, t_l, \label{1} \\
&y_{\text{min}} \le y[t_l] \le y_{\text{max}}, \quad \forall l, t_l, \label{2}\\
&H_{\text{min}} \le H[t_l] \le H_{\text{max}}, \quad \forall l, t_l, \label{3}\\
&| \boldsymbol{w}[t_l+1] - \boldsymbol{w}[t_l] |^2 \leq S_{xy_{\text{max}}}^2, \quad \forall l, t_l, \label{4}\\
&| H[t_l+1] - H[t_l] |^2 \leq S_{H_{\text{max}}}^2, \quad \forall l, t_l.\label{5}
\end{align}
Here, $(.)_{\text{min}}$ and $(.)_{\text{max}}$ denote the bounds for the UAV coordinates, and $S(.)_{\text{max}} = V(.)_{\text{max}} \delta$ represents the maximum distance the UAV can travel in each time frame. Due to limited communication resources (e.g., bandwidth and power), the UAV serves up to $C_{\max}$ users per time slot $t_l$ \cite{10122731}. To model user connections, we define a binary variable $\small{\boldsymbol{J}[t_l]=[J_n^l[t_l]]^{N \times L}}$, where $\small{J_n^l[t_l]=1}$ if user $n$ is connected to the UAV at $t_l$, and $J_n^l[t_l]=0$ otherwise. Similarly, to maintain QoS ($r_{on}^l$), users must remain connected for at least $\tau_n$ slots. This setup imposes the following constraints:
\begin{align}
&{\sum}_{{t_l}\in \mathcal{T}_l} J^l_n[t_l] \geq \tau_n, \quad \forall l,n, \label{6}\\
&{\sum}_{n\in \mathcal{N}_l} J^l_n[t_l] \leq C_{\text{max}}, \quad \forall l, t_l, \label{7}\\
&{\sum}_{l\in \mathcal{L}} J^l_n[t_l] \leq 1, \quad \forall n, t_l.\label{8}
\end{align}
\subsection{Transmission and Channel Modeling}
This work adopts a probabilistic channel model accounting for LoS and NLoS components. LoS probability ($P_{\text{LoS}}$) is a function of user-UAV 3D positions and the surrounding environment \cite{7918510}, given by  
${\text{PLoS}_n^l[t_l] = (1 + b_1 \exp(-b_2(\theta_n^l[t_l] - b_1)))^{-1}}$,  
where $b_1$ and $b_2$ represent environmental conditions. The elevation angle $\theta_n^l[t_l]$ is  
${\theta_n^l[t_l] = \sin^{-1} \left(\frac{H[t_l]}{d_n^l[t_l]}\right)}$,  
where the distance $d_n^l[t_l]$ is  
${d_n^l[t_l] \!\!=\!\! \sqrt{\|\boldsymbol{w}[t_l]\!\! -\! \boldsymbol{Q}_n^l[t_l]\|^2\! +\! H^2[t_l]}}$.  
Based on $\text{PLoS}_n^l[t_l]$, the path loss $\text{PL}_n^l[t_l]$ is  
$\text{PL}_n^l[t_l] = \eta_{(.)} \left(\frac{4\pi f_c d_n^l[t_l]}{c}\right)^2$,  
where $f_c$ is the carrier frequency and $c$ the speed of light. The attenuation factor $\eta_{(.)}$, determined by $\text{PLoS}_n^l[t_l]$, influences $\text{PL}_n^l[t_l]$ with $\eta_{\text{LoS}}$ for  
$\text{PLoS}_n^l[t_l] \ge \phi$ (threshold) and $\eta_{\text{NLoS}}$ otherwise.  
Using this, the effective channel gain with maximum ratio transmission precoding is  
$g_n^l[t_l] = \frac{| h_n^l[t_l](h_n^l[t_l])^H|}{PL_n^l[t_l]}$,  
where $h_n^l[t_l] \in \mathbb{C}^{1 \times M}$ represents small-scale fading. Based on $g_n^l[t_l] $, the SNR for the $n$-th user associated with $l$-th cluster can be expressed as:
 $   \gamma_n^l[t_l]=p_n^l[t_l]g_n^l[t_l]/\sigma^2.$
Where $\sigma^2 = B_n^l[t_l] N_o$, with $B_n^l[t_l]$ as $n$-th user bandwidth, $N_o$ the noise spectral density, and $p_n^l[t_l]$ the transmitted power in cluster $l$ at time slot $t_l$. The achievable data rate for user $n$ is:
\begin{equation}
\label{ratee}
R_n^l=1/T_l{\sum}_{{t_l}\in \mathcal{T}_l}B_n^l[t_l]log_2(1+    \gamma_n^l[t_l]).
\end{equation}
\subsection{Problem Formulation}
\label{PF}
This work aims to enhance the communication system by maximizing the minimum user data rate, a key performance metric that ensures fairness among users. To achieve this, we jointly optimize UAV trajectory, user association, clustering, time slot allocation, flight time, transmit power, and bandwidth. For shorthand, we define $\boldsymbol{B}[t_{l}]=[B_n^l[t_{l}]]^{N \times L}$, $\boldsymbol{P}[t_{l}]=[p_n^l[t_{l}]]^{N \times L}$, and $\boldsymbol{\tau}=[\tau]^{N\times1}$. The resulting joint optimization problem is formulated as follows:
\begin{subequations}
\label{JOP}
\begin{align}
 \max_{\substack{\boldsymbol{J}[{t_l}],\boldsymbol{W}[t_{l}],\boldsymbol{B}[t_{l}],\\\boldsymbol{P}[t_{l}],T_l,\boldsymbol{\tau},L}}& \quad \min_{\forall n}  {\sum}_{l=1}^{L} R_n^l, \\ 
\text{s.t.} &\text{\cref{1,2,3,4,5,6,7,8}}.\notag\\ 
&R_n^l\tau_n\ge \boldsymbol{J}_n^lr_{on}^l, \forall n,l \label{C1}\\
&B_n^l[t_{l}] \le J_n^l[t_l]B^{\text{max}},\forall n, l,t_l\label{C2}\\
&p_n^l[t_{l}] \le J_n^l[t_l]P^{\text{max}},\forall n, l,t_l\label{C2a}\\
& {\sum}_{n\in \mathcal{N}_l}p_n^l[t_{l}]\le P_t^{\text{max}}, \forall l,t_l,\label{C3}\\
&{\sum}_{n\in \mathcal{N}_l}B_n^l[t_{l}]\le B^{\text{max}}, \forall l,t_l,\label{C4}\\
&{\sum}_{l \in \mathcal{L}} T_l \le T, \label{C5}\\
&\boldsymbol{W}[0] = \boldsymbol{W}[T]\label{C6}.
\end{align}
\end{subequations}
where $\boldsymbol{J}_n^l = \sum_{t_l \in \mathcal{T}_l} \boldsymbol{J}_n^l[t_l]$. Constraint \eqref{C1} ensures service quality, requiring user $n$ to receive at least $r_{on}^l$ bits. Constraints \eqref{C2} and \eqref{C2a} allocate bandwidth and power only for users associated with the UAV. To enforce resource limits, \eqref{C3} and \eqref{C4} constrain the maximum transmit power $P_t^{\text{max}}$ and bandwidth $B^{\text{max}}$.  
Additionally, \eqref{C5} restricts the UAV’s flight duration, while \eqref{C6} ensures it returns to its initial position at time $T$. 
\section{Framework for UAV Trajectory, User Association and Resource Allocation} \label{FW} 
The optimization problem \eqref{JOP} is an NP-hard and mixed integer nonlinear program (MINLP) due to the integral nature of the user association variable $\!\!\boldsymbol{J}_n^l[t_l]$, non-linear because of logarithmic rate function, and nonconvex because variable coupling. Moreover, integrating user mobility with physical layer resource allocation further complicates joint UAV trajectory and resource optimization. To enhance tractability and efficiency, we apply the block coordinate descent (BCD) method to decouple the problem into two subproblems: UAV trajectory and user association (Section \ref{TAA}) and joint bandwidth and power allocation (Section \ref{JBPAS}), following the user mobility prediction model in Section \ref{MM}. 
\subsection{User Mobility Prediction Model:}\label{MM}
This section presents a Markov chain-based mobility prediction model \cite{jiang1988bridge} deployed on a UAV, predicting user locations based on prior information from the CCU.
We define a set of $K$ possible states, $\mathcal{S} = \{s_1, s_2, \dots, s_K\}$, including the current state $s_c$, forward/backward states $s_f, s_b$, left/right states $s_l, s_r$, and other directions $\{s_i\!\! \mid \!\!i \!\notin \!{c,\! f,\! b,\! l,\! r}\}$ for each user $n$. The state transitions follow a second-order Markov model, represented by a three-dimensional transition matrix $\Omega = [\Omega_{ijk}]^{K \times K \times K}$, where $\Omega_{ijk}$ denotes the probability of transitioning from state $s_k$ to $s_j$, given $s_i$ and  computed by normalizing the observed transition frequencies $\rho_{ijk}$:
\begin{equation}
\Omega_{ijk} = \frac{\rho_{ijk}}{\sum_{l \in \mathcal{S}} \rho_{ijl}}, \quad {\sum}_{k \in \mathcal{S}} \Omega_{ijk} = 1, \forall i,j \in \mathcal{S},
\end{equation}
where $\sum_{l \in \mathcal{S}} \rho_{ijl}$ represents total transitions from $s_i$ to $s_j$ and subsequently to any state $s_l$, ensuring probabilistic consistency. The initial state distribution $\boldsymbol{\pi}_n[0]$ defines the probability of user $n$ occupying each state at $t=0$:
\begin{equation}
\boldsymbol{\pi}_n[0] = [\pi_{1,n}[0], \pi_{2,n}[0], \dots, \pi_{K,n}[0]],
\end{equation}
where $\pi_{i,n}[0]\!\! = \!\!1$ if $i\! =\! c$ and 0 otherwise. The state distribution evolves as:
\begin{equation}
\label{FLa}
\boldsymbol{\pi}_n[t_l] = \boldsymbol{\pi}_n[t_{l-1}] \cdot \boldsymbol{\Omega}_{\text{prev}, \text{curr}, :}.
\end{equation}
Computing \eqref{FLa} requires $O(K^3)$ operations per transition and is computationally expensive for large $K$. To mitigate this, we exploit the sparsity of $\boldsymbol{\Omega}$, reducing complexity to $O(N_z)$, where $N_z$ is the nonzero count:
\begin{equation}
\label{FL}
\boldsymbol{\pi}_n[t_l] = \boldsymbol{\pi}_n[t_{l-1}] \cdot \text{sparse}(\boldsymbol{\Omega}_{\text{prev}, \text{curr}, :}).
\end{equation}
Using \eqref{FL}, the most probable user state at $t_l$  determined via Maximum A Posteriori (MAP) estimation \cite{gauvain1994maximum}:
\begin{equation}
s_n[t_l] = \arg \max_{k \in \mathcal{S}} \pi_{k,n}[t_l].
\end{equation}
Each state corresponds to a movement vector that updates the user’s position as: 
\begin{equation}
\boldsymbol{Q}_n[t_l] = \boldsymbol{Q}_n[t_{l-1}] + \Delta \boldsymbol{Q}_n[t_l].
\end{equation}
By leveraging sparsity, the prediction model reduces the per-step complexity from \( O(K^3) \) to \( O(N_z) \). The total complexity includes state initialization \( O(N_l K) \), and \( O(N_l T_l N_z) \) for both state updates and MAP estimation, where steps are detailed in Algorithm~\ref{MobilityPrediction}. 
\begin{algorithm2e}
\SetAlgoLined
\KwIn{Initial state $\boldsymbol{\pi}_n[0]$, transition matrix $\boldsymbol{\Omega}$, time slots $T_l$, users $N_l$, states $\mathcal{S}$}
\KwOut{Predicted locations $\boldsymbol{Q}_n[t_l]$}
\For{each user $n$}{
    \For{each time slot $t_l$}{
        Update state distribution: $\boldsymbol{\pi}_n[t_l] \leftarrow \boldsymbol{\pi}_n[t_l-1] \cdot \text{sparse}(\boldsymbol{\Omega})$\;
        Predict location: $\boldsymbol{Q}_n[t_l] \leftarrow \arg \max_{k \in \mathcal{S}} \pi_{k,n}[t_l]$\;
    }
}
\caption{Mobility Prediction Algorithm}
\label{MobilityPrediction}
\end{algorithm2e}
\subsection{UAV Trajectory and User Association:}  
\label{TAA}
Given the users' locations at time slot $t_l$ and fixed radio resources $(\boldsymbol{B}[t_{l}], \boldsymbol{P}[t_{l}])$, the subproblem for UAV trajectory and user association is formulated as follows:
\begin{subequations}
\label{JOP1}
\begin{align}
\!\!\!& \max_{\substack{\boldsymbol{J}[{t_l}],\boldsymbol{W}[t_{l}],T_l,\boldsymbol{\tau},L}} \quad \min_{\forall n} {\sum}_{l=1}^{L} R_n^l, \label{OP1}\\ 
\text{s.t.} &\text{\cref{1,2,3,4,5,6,7,8,C1,C2,C5,C6}}.\notag 
\end{align} 
\end{subequations} 
Moreover, joint optimization of UAV trajectory $(\boldsymbol{W}[t_{l}])$, user association $(\boldsymbol{J}[{t_l}])$, clustering $(L)$, serving time $(T_l)$, and user connectivity $(\boldsymbol{\tau})$ is still computationally challenging due to their mutual coupling and high dimensionality. To tackle this, we propose a heuristic approach to efficiently determine the number of clusters, serving time, and user connectivity. The $L$  clusters are determined using Shannon’s capacity bound, which estimates the average spectral efficiency: 
\begin{equation}
\label{ESE}
\Lambda = \mathbb{E}\left[\log_2\left(1 + \frac{P_t^{\max}}{f_c \sigma^2} \left(\frac{\left|\mathbf{h}^H \mathbf{h}\right|^2}{PL}\right)\right)\right].
\small
\end{equation}
Assuming LoS with $\text{PLoS} \approx 1$, the \eqref{ESE} provides an upper bound on the achievable data rate: $R^{\max} = B \times \Lambda$. To satisfy the minimum required data rate $r_{on}$ in constraint \eqref{C1}, the user connectivity time is given by: $\tau = \left\lceil \frac{r_{on}}{R^{\max}} \right\rceil.$
From this, the UAV’s maximum user capacity per time slot is: $C^{\max} = {\left\lfloor \frac{\tau\times R^{\max}}{r_{on} } \right\rfloor}.$ Using $C^{\max}$ as an estimate of cluster size, users are grouped into $L = \left\lceil \frac{N}{C^{\max}} \right\rceil$ clusters via K-means. With clusters and connectivity times established, the next step is to optimize the UAV’s trajectory and serving schedule. The problem remains nonlinear and non-convex due to the fractional term $T_l$ in \eqref{OP1} and the dynamic nature of user mobility. To overcome this, we adopt an iterative strategy where the UAV first selects the nearest cluster:
\begin{equation}
\label{18}
l_{\min} = \arg \min_{l \in \{1, \ldots, L\}} \|\boldsymbol{W}[t_l] - \boldsymbol{Q}^l[t_l]\|.
\end{equation}
During $T_l$, the UAV tracks and serves users in the $l$-th cluster, reducing computational complexity since $N_l \ll N$. The serving time for each cluster is: $T_l = f(N_l, D_l),$ Where $N_l$ is the number of users in the cluster, and $D_l$ is the UAV’s travel distance to its centroid, ensuring clusters with higher user densities receive more service time. Once $T_l$ elapses, the UAV updates its position, gathers user locations, and dynamically re-clusters the remaining users into $L' \leq L$ groups. This process iterates until all users are served, balancing user association, trajectory optimization, and computational efficiency. To enhance tractability, we introduce the auxiliary variable: $\Gamma = \min R_n^l, \quad \forall n,$, which allows us to decouple user association and trajectory optimization into separate subproblems. This structured approach simplifies problem-solving and enhances resource utilization and operational efficiency in UAV-assisted networks.
\subsubsection{Optimizing User Association}
Given the $\boldsymbol{W}[t_{l}]$, the user association subproblem can be formulated as:  
\begin{subequations}
\label{JOP3}
\begin{align}
& \max_{\substack{\boldsymbol{J}[{t_l}],\Gamma}} \; \Gamma, \text{s.t.} \; R_n^l \ge \Gamma, \; \Gamma\tau_n \ge  r_{on}^l, \forall n \in \mathcal{N}_l\label{C1a} \\
&\text{\cref{6,7,8,C2}}. \notag
\end{align}
\end{subequations}
The optimization problem \eqref{JOP3} is an integer linear and convex problem. This can be solved using standard methods like integer linear programming (ILP).
\subsubsection{UAV Trajectory Optimization}
Following that, under the given user association matrix, the subproblem for UAV trajectory optimization can be expressed as: 
\begin{subequations}
\label{JOP2}
\begin{align}
\!\!\!& \max_{\substack{\boldsymbol{W}[t_{l}],\Gamma}} \quad \Gamma, \label{OP2}\\
&R_n^l \ge \Gamma, \forall n \in \mathcal{N}_l\label{C1c} \\
\text{s.t.} &\text{\cref{1,2,3,4,5,C1a,C6}}.\notag
\end{align}
\end{subequations}
The objective function \eqref{OP2} in \eqref{JOP2} is convex, but the term \( R_n^l \)  defined in \eqref{ratee} remains non-convex due to the \( \log(\cdot) \) function. To simplify and enhance tractability, we define \(\!\!PL_o \!\!=\!\! \frac{1}{\eta_{(.)} \!\left( \frac{4\pi f_c}{c} \right)^2 \sigma^2} \) and rewrite \eqref{ratee} as:  
\begin{equation}
\label{ratee1}
R_n^l = \sum_{{t_l} \in \mathcal{T}_l} \frac{B_n^l[t_l]}{T_l} f_n^l(\boldsymbol{w}[t_l], H[t_l]);
\end{equation}
\begin{equation}
\label{RR}
\!\!\!\!\!\!\!f_n^l(\boldsymbol{w}[t_l],\! H[t_l])\!\! =\! \log_2 \!\!\left( \!\!1 \!+\! \frac{p_n^l[t_l] PL_o} {\|\!\boldsymbol{w}[t_l]\!\! -\! \boldsymbol{Q}_n^l[t_l]\|^2 \!\!+\!\! H^2[t_l]} \!\!\right)\!.
\end{equation}
To simplify the non-convex problem, we apply a first-order Taylor expansion around the point $(\boldsymbol{w}_0, H_0)$. For the shorthand notation, we introduce: $\mathcal{C}^l_{1,n}[t_l] = \|\boldsymbol{w}_0[t_l] - \boldsymbol{Q}_n^l[t_l]\|^2 + H_0^2[t_l],$
$\mathcal{C}^l_{2,n} [t_l] = p_n^l[t_l] PL_o, $
$\mathcal{C}_3 = \ln(2)$ and expressed the Taylor expression as:  
\begin{align}
\label{TEX}
&f(\|\boldsymbol{w}[t_l]\|^2, H^2[t_l]) \approx f(\|\boldsymbol{w}_0[t_l]\|^2, H_0^2[t_l]) \notag \\
&+ \left.\frac{\partial f} {\partial \|\boldsymbol{w}\|^2} \right|_{(\|\boldsymbol{w}_0[t_l]\|^2, H_0^2[t_l])} (\|\boldsymbol{w}[t_l]\|^2 - \|\boldsymbol{w}_0[t_l]\|^2) \notag \\ 
&+ \left.\frac{\partial f}{\partial H^2}\right|_{(\|\boldsymbol{w}_0[t_l]\|^2, H_0^2[t_l])} (H^2[t_l] - H_0^2[t_l]).
\small
\end{align}
The partial derivatives of \eqref{RR} with respect to $\|\boldsymbol{w}\|^2$ and $H^2$ are given by:  
\begin{equation}
\label{Derivative}
\small
\!\!\!\!\!\!\left.\frac{\partial f}{\partial ||\boldsymbol{w}||^2}\right|_{(\cdot)} \!\!\!\!\!\!= \!\!\left.\frac{\partial f}{\partial H^2}\right|_{(\cdot)}\!\!\!\!\!\! = \!\!- \frac{\mathcal{C}^l_{2,n}[t_l]}{\mathcal{C}_3 \mathcal{C}^l_{1,n}[t_l] (\mathcal{C}^l_{1,n}[t_l] + \mathcal{C}^l_{1,n}[t_l])}=\!\nabla\!{f}^l_{n}[t_l].
\small
\end{equation}
where $(\cdot)={(||\boldsymbol{w}_0[t_l]||^2, H_0^2[t_l])} $. Substituting \eqref{Derivative} into the \eqref{TEX} gives the first-order approximation of \eqref{RR} as shown \eqref{FTA}.
\begin{figure*}[htb!]
\begin{align}
\label{FTA}
\!f_n^l(\|\boldsymbol{w}[t_l]\|^2, H^2[t_l]) \!\!&\approx\!\! f(\|\boldsymbol{w}_0[t_l]\|^2\!\!, \!H_0^2[t_l])\! +\!\!\nabla\!{f}^l_{n}[t_l](\|\boldsymbol{w}_0[t_l]\|^2\!\!,\! H_0^2[t_l])\! \cdot\! (\|\boldsymbol{w}[t_l]\|^2\!\! -\! \|\boldsymbol{w}_0[t_l]\|^2\!\! +\!\! H^2[t_l]\!\! - \!\!H_0^2[t_l])
\end{align}
\hrule
\end{figure*}
Following that, by substituting  \eqref{FTA} into the \eqref{ratee1}, we obtain the lower bound $(Rlb_n^l)$ for $R_n^l$ as:    
\begin{equation}
R_n^l \approx Rlb_n^l\geq \sum_{{t_l} \in \mathcal{T}_l} \frac{B_n^l[t_l]}{T_l} \left[f_n^l(\|\boldsymbol{w}[t_l]\|^2, H^2[t_l]) \right]  
\end{equation}
This lower bound simplifies the \eqref{JOP2}, enabling tractable optimization. The joint problem is then formulated as:    
\begin{subequations}
\label{JOP4}
\begin{align}
\!\!\!& \max_{\substack{\boldsymbol{W}[t_{l}],\Gamma}} \quad \Gamma, \text{s.t.}\; Rlb_n^l \ge \Gamma, \forall n \in \mathcal{N}_l\\
\text{s.t.} &\text{\cref{1,2,3,4,5,C1a,C6}}.\notag       
\end{align} 
\end{subequations}
The optimization problem \eqref{JOP4} is convex, enabling a (suboptimal) solution for \eqref{JOP3} using CVX. To bridge the gap between optimal and suboptimal solutions, the UAV’s initial trajectory is set according to \cite{8247211} for $N_l$ users in the $l$-th cluster, while adhering to horizontal ($V_{xy_{\text{max}}}$) and vertical ($V_{H_{\text{max}}}$) speed constraints in \eqref{4} and \eqref{5}. The UAV follows a circular trajectory in 3D space at a constant speed $V = \sqrt{V_{xy}^2 + V_{H}^2}$.  The trajectory’s center and radius are defined as $\mathbf{c}_{\text{trj}} = [x_{\text{trj}}, y_{\text{trj}}, z_{\text{trj}}]^T$ and $\rho_{\text{trj}}$, respectively, where $2\pi \rho_{\text{trj}} = V T_l$ for any period $T_l$. The geometric center of users in cluster $l$ is: $\mathbf{c}_{\text{g}} = \frac{\sum_{n=1}^{N_l} \mathbf{Q}_n[t_l]}{N_l}.$ The radius $\rho_{\text{u}}$ represents the farthest user from $\mathbf{c}_{\text{g}}$, given by: $\rho_{\text{u}} = \max_{n \in N_l} |\mathbf{Q}_n[t_l] - \mathbf{c}_{\text{g}}|.$ Using a circle packing scheme (CP), the trajectory’s center $\mathbf{c}_{\text{trj}}$ and radius $\rho_{\text{cp}}$ are determined. To ensure feasibility, the trajectory radius is capped by: $\rho_{\text{trj}} = \min(\rho_{\max}, \frac{\rho_{\text{cp}}}{2}), \quad \text{where } \rho_{\max} = \frac{V_{\max} T_l}{2\pi}.$ Thus, the initial trajectory at time slot $t_l$ is:
\begin{equation}
\hat{\mathbf{W}} [t_l] = \begin{bmatrix}
x_{\text{trj}} + \rho_{\text{trj}} \cos \theta_{t_l} \\
y_{\text{trj}} + \rho_{\text{trj}} \sin \theta_{t_l} \\
z_{\text{trj}}
\end{bmatrix},  \theta_{t_l} \triangleq 2\pi \frac{(t_l-1)}{T_l-1}.
\end{equation}
This ensures the UAV trajectory adheres to coverage and speed constraints, as outlined in Algorithm \ref{Algo1}  
\begin{algorithm2e}[htb!]
\SetAlgoLined
\KwIn{ $\boldsymbol{Q}_n[t_l]$,$\boldsymbol{B}$, $\boldsymbol{P}$}
\KwOut{Optimized  $\boldsymbol{W}[t_l]$,  $\boldsymbol{J}[t_l]$}
Solve user association: $\boldsymbol{J}_i^* \leftarrow $ solve\eqref{JOP3}\;
Optimize UAV trajectory: $\boldsymbol{W}_i^* \leftarrow$ solve  \eqref{JOP4}\;
\caption{\small{Joint UAV Trajectory and Association}}
\label{Algo1}
\end{algorithm2e}

\vspace{-2mm}
   
\subsection{Joint Bandwidth and Power Allocation}
\label{JBPAS}
Under the given trajectory and user association matrix, the joint optimization problem for bandwidth and power allocation is as follows:
\begin{subequations}
\label{JBPA}
\begin{align}
& \max_{\substack{\boldsymbol{B}[t_{l}],\boldsymbol{P}[t_{l}],\Gamma}} \quad \Gamma, \\
\text{s.t.} & \frac{1}{T_l}\sum_{t_l \in \mathcal{T}_l} B_n^l[t_l] \log_2(1+ \gamma_n^l[t_l]) \ge \Gamma, \forall n \in \mathcal{N}_l \label{C1b1} \\ 
& \Gamma \tau_n \ge \boldsymbol{J}_n^l r_{on}^l, \forall n \in \mathcal{N}_l \label{C1a1} \\
& \text{\cref{C2,C3,C4}}. \notag
\end{align}
\end{subequations} 
where \( \gamma_n^l[t_l] \!=\! \frac{p_n^l[t_l] g_n^l[t_l]}{B_n^l[t_l] N_o} \), the problem \eqref{JBPA} remains nonconvex due to constraint \eqref{C1b1}. We introduce auxiliary variables \( \boldsymbol{\Psi}[t_l] = [\Psi_n^l[t_l]]^{N \times L} \) and rewrite \eqref{JBPA} as:
\begin{subequations} 
\label{JBPA1}
\begin{align}
& \max_{\substack{\boldsymbol{B}[t_{l}],\boldsymbol{P}[t_{l}],\Gamma}} \quad \Gamma, \\
\text{s.t.}\! & \frac{1}{T_l}\!\!{\sum}_{t_l \in \mathcal{T}_l} B_n^l[t_l] \log_2(1+ \Psi_n^l[t_l]) \!\!\ge \!\!\Gamma, \!\forall n \in \mathcal{N}_l \label{C1b2} \\ 
& \frac{p_n^l[t_l] g_n^l[t_l]}{B_n^l[t_l] N_o} \ge \Psi_n^l[t_l], \forall n \in \mathcal{N}_l, t_l \in \mathcal{T}_l, \label{C2b2} \\
& \text{\cref{C2,C3,C4,C1a1}}. \notag
\end{align}
\end{subequations}
The problem \eqref{JBPA1} is equivalent to \eqref{JBPA}. Demonstrate this equivalence by showing that the inequalities in constraints \eqref{C1b2} and \eqref{C2b2} are satisfied. The challenge lies in solving \eqref{C1b2} and \eqref{C2b2}. To address this, we introduce the slack variable $\boldsymbol{\psi}[t_{l}]=[\psi_n^l[t_{l}]]^{N \times L}$  and reformulate constraint \eqref{C1b2} as:
\begin{align}
& \log_2(1+ \Psi_n^l[t_l]) \ge \psi_n^l[t_l], \forall n \in \mathcal{N}_l, t_l \in \mathcal{T}_l. \label{C1b3} \\
& \frac{1}{T_l}{\sum}_{t_l \in \mathcal{T}_l} B_n^l[t_l] \psi_n^l[t_l] \ge \Gamma, \forall n \in \mathcal{N}_l \label{C1b4}.
\end{align}
The constraint \eqref{C1b3} is convex in nature. To deal with constraint \eqref{C1b4}, we first reformulate it into a difference-of-convex presentation and then apply the first-order approximation \cite{vu2021dynamic}.
Given the feasible solution $(\hat{B}_n^l[t_l], \hat{\psi}_n^l[t_l])$ from the current iteration, constraint \eqref{C1b4} is approximated in the next iteration as:
\begin{align}
\label{33}
& \sum_{t_l \in \mathcal{T}_l} 2(\hat{B}_n^l[t_l] + \hat{\psi}_n^l[t_l])(B_n^l[t_l] + \psi_n^l[t_l]) \\
& \!\!\!\! -\!\!\!\! \sum_{t_l \in \mathcal{T}_l} (\hat{B}_n^l[t_l]\! +\! \hat{\psi}_n^l[t_l])^2 \!\ge \!\!\!\!\sum_{t_l \in \mathcal{T}_l} (B_n^l[t_l]^2 \!+ \!\psi_n^l[t_l]^2)\! + \!2T_l \Gamma \notag.
\end{align}
To address constraint \eqref{C2b2}, we rewrite it as:
\begin{equation}
\label{Trf}
p_n^l[t_l] g_n^l[t_l] N^{-1}_o \ge \Psi_n^l[t_l] B_n^l[t_l].
\end{equation}
Since the left side of \eqref{Trf} is linear and convex, while the right side is neither, we apply a first-order approximation to the nonconvex side. Given the feasible solution $(\hat{B}_n^l[t_l], \hat{\Psi}_n^l[t_l])$, approximate constraint \eqref{C2b2} as:
\begin{equation}
\Psi_n^l[t_l] B_n^l[t_l] \approx \hat{B}_n^l[t_l] \Psi_n^l[t_l] + \hat{\Psi}_n^l[t_l] B_n^l[t_l].
\end{equation}
\begin{equation}
\label{ftrac}
\frac{g_n^l[t_l] p_n^l[t_l]}{N_o} \ge \hat{B}_n^l[t_l] \Psi_n^l[t_l] + \hat{\Psi}_n^l[t_l] B_n^l[t_l].
\end{equation}
Using \eqref{ftrac} and \eqref{33}, the inner approximations of \eqref{Trf} and \eqref{C1b4}, respectively, we approximate the \eqref{JBPA} as:
\begin{subequations}
\label{FTAA}
\begin{align}
& \max_{\substack{\boldsymbol{B}[t_{l}],\boldsymbol{P}[t_{l}],\Gamma,\boldsymbol{\psi}[t_{l}],\boldsymbol{\Psi}[t_{l}]}} \quad \Gamma, \\
& \text{\cref{C2,C3,C4,C1a1,C1b3,ftrac,33}}. \notag
\end{align}
\end{subequations}
We observe that problem \eqref{FTAA} is convex and can be efficiently solved using convex optimization tools such as CVX. Since its solution satisfies \cref{C1b1,C1a1,C2,C3,C4}, it provides a (sub)optimal solution for \eqref{JBPA}. The performance gap between this (sub)optimal solution and the global optimum of \eqref{JBPA} primarily depends on the initialization of $\hat{\boldsymbol{\Psi}}, \hat{\boldsymbol{\psi}}, \hat{\boldsymbol{B}}$, and $\hat{\boldsymbol{P}}$. To address this, we propose an iterative method that refines the solution through a series of convex subproblems, as detailed in Algorithm \ref{Algo2}.
%
    \begin{figure}[h]
	\centering
	\includegraphics[width=0.85\columnwidth]{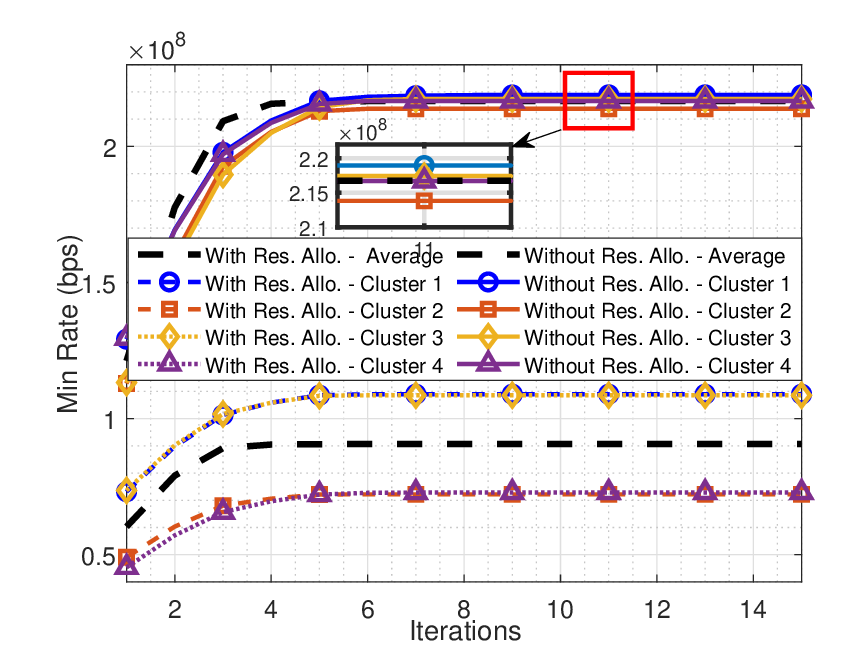}
	\caption{Convergence Analysis}
	\label{fig:convergance}
\end{figure}
\begin{figure*}[h]
	\centering
	\begin{subfigure}[b]{0.34\linewidth}
		\centering
		\includegraphics[width=\linewidth]{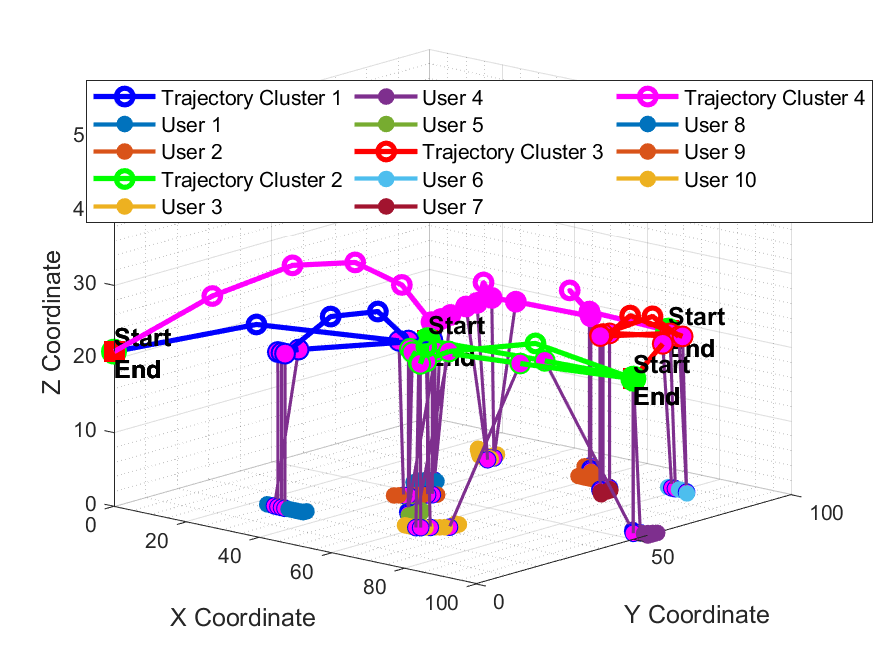}
		\caption{{\small 3D UAV flight trajectory.}}
		\label{fig:Trajectory}
	\end{subfigure}%
	\begin{subfigure}[b]{0.34\linewidth}
		\centering
		\includegraphics[width=\linewidth]{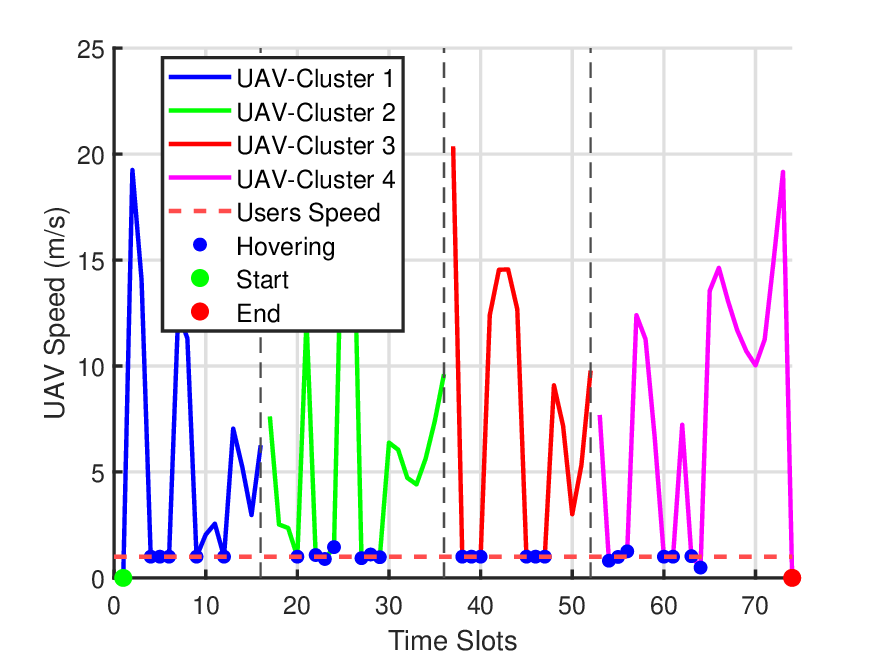}
		\caption{UAV speed variations.}
		\label{fig:SpeedGraph}
	\end{subfigure}%
	\begin{subfigure}[b]{0.34\linewidth}
		\centering
		\includegraphics[width=\linewidth]{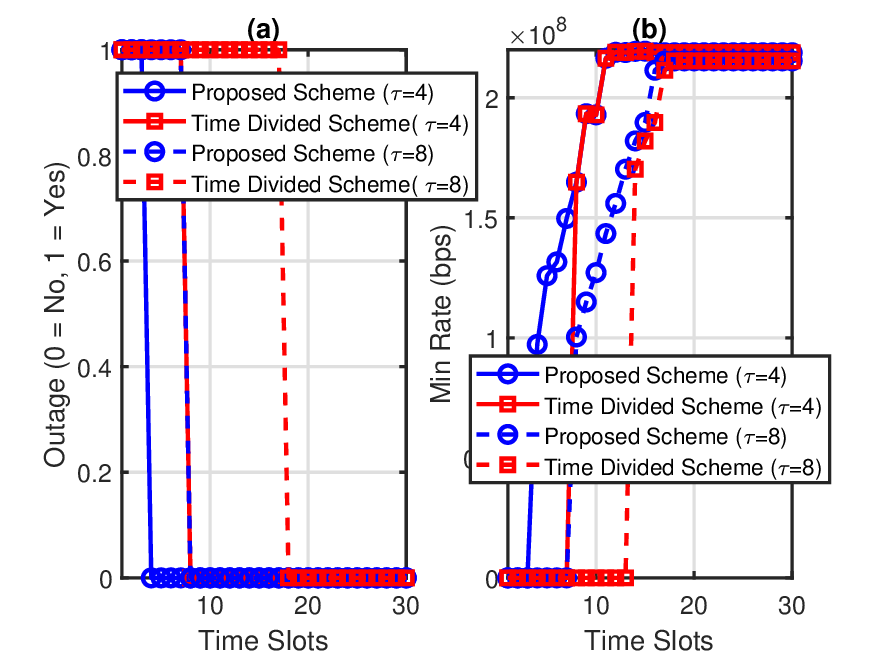}
		\caption{{ Proposed vs.Time Dividend \cite{9678115}.}}
		\label{fig:OutageGraph}
	\end{subfigure} 
	\caption{UAV trajectories and user associations with optimal resource allocation.}
	\label{fig:TandUserAssociation}
\end{figure*}
\begin{algorithm2e}
\SetAlgoLined
\KwIn{Variables $\hat{\boldsymbol{\Psi}}$, $\hat{\boldsymbol{\psi}}\!$, $\hat{\boldsymbol{B}}$, $\hat{\boldsymbol{P}}$, $\boldsymbol{Q}_n$, $\boldsymbol{W}$, $\boldsymbol{J}$, $\forall t_l \!\in \!\!\mathcal{T}$, $\forall n \!\in\! \mathcal{N}$, $\mathcal{F}_{\text{prev}} \leftarrow 0$, $P_t^{max}$, $B_t^{max}$, $r_{on}^l$, maximum iterations $\!I_{\max}$, tolerance $\!\epsilon$.}
\KwOut{${\boldsymbol{B}}, {\boldsymbol{P}}, \forall t_l \in \mathcal{T}, \forall n \in \mathcal{N}_l$}
\BlankLine
    $[\boldsymbol{B}^*_i, \boldsymbol{P}^*_i, \Gamma^*_i, \boldsymbol{\psi}^*_i, \boldsymbol{\Psi}^*_i] \leftarrow$ solve \eqref{FTAA} using CVX\;
Update: $[\hat{\boldsymbol{\Psi}},\hat{\boldsymbol{\psi}},\hat{\boldsymbol{B}},\hat{\boldsymbol{P}}]$ $\leftarrow [\boldsymbol{\Psi}^*_i, \boldsymbol{\psi}^*_i,\boldsymbol{B}^*_i, \boldsymbol{P}^*_i].$
\caption{\!{Joint Resource Allocation}}
\label{Algo2}
\end{algorithm2e}

\begin{figure}[h]
	\centering
	\includegraphics[width=0.85\linewidth]{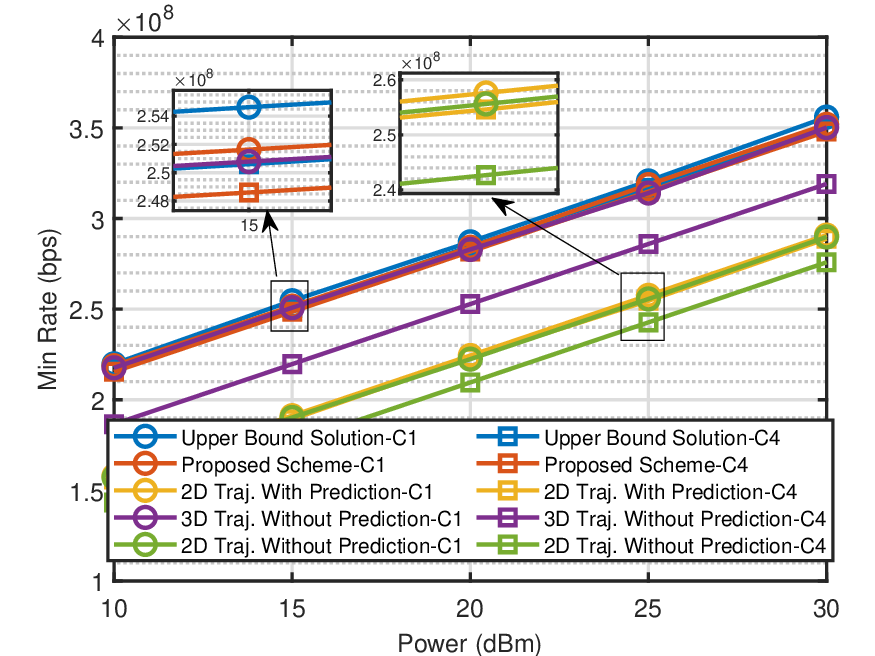}
	\caption{Performance Across Different  Schemes. Legend abbreviations: C1 Cluster 1, C4 Cluster 4.}
	\label{fig:R7}
\end{figure}
\vspace{-5mm}
\subsection{Joint Algorithm Design and Complexity Analysis}  
\label{CCAA}
The iterative algorithm solves \eqref{JOP} to obtain a (sub)optimal solution by determining the minimum user connectivity time $\boldsymbol{\tau}$, UAV flight time $T_l$, number of clusters $L$, and user partitioning. User locations $\boldsymbol{Q}_n[t_l]$ are first predicted using a mobility model. Then, Algorithm~\ref{Algo1} jointly optimizes user association and UAV trajectory with complexity $C1 = \mathcal{O}(2^{\mathcal{N}_l} + Z^{3.5} \log(1/\epsilon))$, where \( Z \) is the number of decision variables. Algorithm~\ref{Algo2} allocates bandwidth and power by solving the convex problem \eqref{FTAA} via the interior point method, with complexity $C2 = \mathcal{O}((4N_lT_l + 1)^3)$. The joint solution at iteration $i$ is $\boldsymbol{\Pi}_i^*[t_l] = [\boldsymbol{B}_i^*, \boldsymbol{P}_i^*, \boldsymbol{Q}_i^*, \boldsymbol{W}_i^*, \boldsymbol{J}_i^*]$, and the objective is $\mathcal{F}_{\text{current}} = \min_{\forall n} R_n(\boldsymbol{\Pi}_i^*[t_l], \forall t_l \in \mathcal{T}_l)$. The solution is refined until $|\mathcal{F}_{\text{prev}} - \mathcal{F}_{\text{current}}| < \epsilon$. Since each subproblem’s feasible region is a subset of the original joint problem, and each iteration yields non-decreasing objective values \cite[Sec. III-A]{vu2021dynamic}, convergence is guaranteed. The overall complexity is \( \mathcal{O}(L I_{\max}(C1 + C2)) \), where \( I_{\max} \) is the number of iterations per cluster.
\vspace{-2mm}
\section{Numerical Results} 
This section presents numerical results from MATLAB simulations over independent channel realizations. $10$ mobile users are uniformly distributed across a $100\times100$m area, with constant movement simulated to mimic real-world mobility. The UAV operates under a dynamic 3D trajectory, clustering users and serving them sequentially based on predicted locations. Key simulation parameters include a total flight time of $210$ s, a transmit power of $10$ dBm, a maximum bandwidth of $20$MHz, a frequency of $900$ MHz, a noise power spectral density of $-168$ dBm/Hz, and altitude constraints of $21-100$ meters. The UAV's vertical and horizontal movement is limited to $15$ m and $30$ m, respectively. Performance is compared against five benchmarks: (i) an Upper Bound Solution with optimal 3D trajectory/resource allocation using perfect user location knowledge \cite{9757149},(ii) the Proposed Scheme without prediction (using current locations only), (iii) the Proposed Scheme with fixed communication resources \cite{9757149,9473504}, (iv) the Time Dividend Approach allocating one user per slot with equal resources \cite{9678115}, and (v)2D Trajectory designs with and without prediction.\par

Fig.~\ref{fig:convergance} compares the convergence of the proposed scheme with/without resource optimization. Both cases converge within 10 iterations (one curve per cluster). The optimized scheme achieves convergence by iteration 8 with minimal deviation (0.67\%), whereas the non-optimized scheme converges faster (iteration 4) but exhibits significantly higher deviation (20\%). The dashed average line underscores resource optimization’s role in ensuring uniformity, highlighting its necessity alongside trajectory design. \par

Fig.~\ref{fig:TandUserAssociation} illustrates the UAV's 3D trajectory and user association, optimizing flight time and resource allocation. Time slots are dynamically assigned (\(C_{\text{max}} = 3\), \(\tau = 4\)) to limit cluster size while ensuring users are served for \(\tau\) slots. The UAV adapts its trajectory Fig. \ref{fig:Trajectory}, re-clustering users after each service session, and adjusts speed as per users Fig.\ref{fig:SpeedGraph} for seamless tracking. Simulations with constrained flight time (\(T_{l,s} < N_l \times \tau\)) evaluate outage probability Fig. \ref{fig:OutageGraph}a and minimum rate Fig. \ref{fig:OutageGraph}b. For \(\tau = 4\) and \(8\), our scheme achieves zero outages 10\% and 30\% faster, respectively, than the Time Dividend Approach, demonstrating the importance of dynamic UAV-user association and trajectory design. \par
Fig. \ref{fig:R7} compares minimum rates across benchmark schemes under transmit powers ranging from 10~dBm to 30~dBm. The proposed scheme with sparse second-order prediction achieves near-upper-bound performance (\(\epsilon \approx\) upper-bound) for both clusters C1 (nearest users) and C4 (farthest users). Notably, the 3D trajectory design \textit{without prediction} matches proposed scheme performance for C1 but lags by 15.51\% for C4, underscoring the prediction model's critical role for distant users. Crucially, the 3D non-prediction design still outperforms 2D-with-prediction trajectories, while 2D designs gain significant improvement from prediction integration. These results demonstrate that 3D trajectory optimization with prediction ensures robust performance across user distributions, particularly for edge-of-coverage scenarios where conventional 2D designs falter.  
\vspace{-4mm}
\section{Conclusion}  
\label{sec:conclusion}  
This work presents a UAV communication framework that combines 3D trajectory optimization with user mobility prediction to improve QoS in dynamic networks. The results show that the sparse second-order prediction model delivers near-optimal performance (\(\epsilon \approx 0.67\%\)) and increases rates for distant users by 16\% compared to the non-predictive 3D designs. Furthermore, 3D trajectory optimization consistently outperforms 2D designs, even without prediction. The integration of adaptive UAV speed control and heuristic clustering (\(C_{\text{max}} = 3, \tau = 4\)) leads to a 10--30\% faster reduction in outages compared to time-division benchmarks. Moreover, simulations conducted under constrained flight time (\(T_{l,s} < N_l \times \tau\)) demonstrate the robustness of the framework, making it particularly suitable for edge-of-coverage scenarios. 

\section*{Acknowledgment}
This research was funded by the Luxembourg National Research Fund (FNR) under grant INTER/MOBILITY/2023/IS/18014377/MCR and project RUTINE (ref. C22/IS/17220888).
\bibliographystyle{IEEEtran}
\bibliography{ReferenceBibFile.bib}
\small
\end{document}